\documentclass[10pt,aps,prd,showpacs,amsmath,amssymb,nofootinbib,twocolumn,longbibliography,superscriptaddress]{revtex4-1}

\usepackage[english]{babel}				
\usepackage[mathscr]{eucal} 			
\usepackage{mathtools}					
\usepackage[ddmmyyyy]{datetime}    		
\usepackage{enumitem}					
\usepackage[dvipsnames]{xcolor}			

\usepackage[colorlinks=true,
            citecolor=ForestGreen,
            linkcolor=red,
            filecolor=cyan,
            urlcolor=magenta,
            backref=false]{hyperref}

\newcommand{\ts}[1]{{\boldsymbol{#1}}}
\newcommand{\dd}{\mbox{d}}
\newcommand{\lap}{\bigtriangleup}

\AtBeginDocument{%
    \newwrite\bibnotes
    \def\bibnotesext{Notes.bib}
    \immediate\openout\bibnotes=\jobname\bibnotesext
    \immediate\write\bibnotes{@CONTROL{REVTEX41Control}}
    \immediate\write\bibnotes{@CONTROL{%
    apsrev41Control,author="08",editor="1",pages="1",title="0",year="1"}}
     \if@filesw
     \immediate\write\@auxout{\string\citation{apsrev41Control}}%
    \fi
}%

\begin{document}


\title{Non-locality and gravitoelectromagnetic duality}

\author{Jens Boos}
\email{jboos@wm.edu}
\affiliation{High Energy Theory Group, Department of Physics, William \& Mary, Williamsburg, VA 23187-8795, United States}

\author{Ivan Kol\'a\v{r}}
\email{i.kolar@rug.nl}
\affiliation{Van Swinderen Institute, University of Groningen, 9747 AG, Groningen, The Netherlands}

\date{\today}

\begin{abstract}
The weak-field Schwarzschild and NUT solutions of general relativity are gravitoelectromagnetically dual to each other, except on the positive $z$-axis. The presence of non-locality weakens this duality and violates it within a smeared region around the positive $z$-axis, whose typical transverse size is given by the scale of non-locality. We restore an exact non-local gravitoelectromagnetic duality everywhere via a manifestly dual modification of the linearized non-local field equations. In the limit of vanishing non-locality we recover the well-known results from weak-field general relativity.
\end{abstract}

\maketitle


\section{Introduction}

Dualities are useful since they reveal seemingly independent concepts to be two sides of the same coin. For example, in four-dimensional spacetime it is possible to convert the electric field of a point charge into the magnetic field of a monopole by a duality transformation (``electromagnetic duality''). There are also dualities between different theories: as it is well known, the weak-field limit of general relativity admits a description closely mimicking Maxwell's equations. This in turn allows one to relate the gravitational field of certain point-like sources to that of spinning sources (``gravitoelectromagnetic duality'').

Gravitoelectromagnetism (``GEM'' in what follows) has a long history. After the early work on the structure of four-dimensional curvature by Bach \cite{Bach:1921}, Rainich \cite{Rainich:1925}, Einstein \cite{Einstein:1927}, Lanczos \cite{Lanczos:1938} (see also Ref.~\cite{Lanczos:1962}) as well as Ruse \cite{Ruse:1944} in the 1920s-1940s, it was Matte \cite{Matte:1953} and Bel \cite{Bel:1962,Bel:2000} who sought to express the description of gravitational waves in a language similar to vacuum electrodynamics. Nowadays, GEM is an established field with many fruitful applications and extensions, and we refer to Refs.~\cite{Maartens:1997fg,Mashhoon:1998tt,Senovilla:2000zp,Costa:2012cw} for additional details, and to Refs.~\cite{Henneaux:2004jw,Bunster:2006rt,Barnich:2008ts,Argurio:2009xr,Castillo1996GravitomagneticMI,Boos:2014hua} for detailed discussions of GEM duality.

In the present paper we would like to understand how fundamental non-locality affects the duality properties of gravitational fields. In particular, we focus our attention on the class of \emph{infinite-derivative} non-local gravity \cite{Modesto:2011kw,Biswas:2011ar} that is under active investigation. While a few exact solutions have been found in the context of gravitational waves \cite{Kilicarslan:2019njc,Dengiz:2020xbu} and cosmology \cite{Biswas:2005qr,Biswas:2010zk}, the complexity of the non-local gravitational field equations has so far prohibited a deeper study of the non-linear regime; a notable exception is the recent work on almost universal spacetimes \cite{Kolar:2021rfl}. At the weak-field level, however, a plethora of solutions has been constructed in the past years \cite{Frolov:2015bia,Frolov:2015bta,Frolov:2015usa,Frolov:2016xhq,Edholm:2016hbt,Boos:2018bxf,Buoninfante:2018rlq,Buoninfante:2018stt,Boos:2020kgj,Boos:2020ccj,Kolar:2020bpo,Kolar:2021oba}. The common feature of these linearized solutions lies in two main aspects:
\begin{itemize}
\item[(i)] At the location of sharp $\delta$-shaped sources, such as point particles, strings, or branes, the gravitational field is smoothed out and manifestly regular.\\[-1.4\baselineskip]
\item[(ii)] At distances much larger than the scale of non-locality $\ell$, the solutions typically approach the solution encountered in local theory.
\end{itemize}
For this reason we expect that GEM duality should be asymptotically recovered in non-local theories. However, at small distance scales this may not be the case, and this paper is devoted to a test of this hypothesis.

We focus our attention on the well-known GEM duality between the Schwarzschild solution \cite{Droste:1916,Schwarzschild:1916}---which can be regarded as the gravitational field of a point particle---and the somewhat more enigmatic Taub--NUT solution \cite{Taub:1950ez,Newman:1963yy}---which may be interpreted as a semi-infinite rotating string \cite{Misner:1963fr}, but see also the related discussion in Refs.~\cite{Dowker:1967zz,bonnor_1969,sackfield_1971,Bonnor1992,Manko:2005nm,griffiths_podolsky_2009,Luna:2015paa,Clement:2015cxa,Gera:2019ebe,Kubiznak:2019yiu,Bordo:2019tyh,Kolar:2019gzy}. Here, the Schwarzschild solution serves as the archetypical gravitoelectric monopole, whereas the Taub--NUT solution plays the role of the gravitomagnetic monopole. Using their weak-field approximations in non-local infinite-derivative gravity we ask: are these geometries still dual to one another?

This paper is organized as follows: In Sec.~\ref{sec:2} we briefly introduce the framework for weak-field infinite-derivative gravity and discuss the role of the Weyl tensor and Ricci tensor in such theories. In Sec.~\ref{sec:3} we introduce GEM quantities for stationary spacetimes, as well as the weak-field Schwarzschild and NUT solutions in infinite-derivative gravity. Sec.~\ref{sec:4} is devoted to a study of the putative GEM duality between the two solutions. Evaluating the electric and magnetic parts of the Weyl curvature we show that an exact GEM duality is spoiled in the presence of non-locality and becomes exact everywhere except on the positive $z$-axis when non-locality vanishes. The duality can be made exact in the local theory, and in the final part of Sec.~\ref{sec:4} we prove that this remains true in the non-local case, and propose a manifestly self-dual non-local model. Therein, any two solutions that are dual in the local model are mapped into dual non-local solutions, and this duality is applicable to a wide range of stationary non-local infinite-derivative gravity theories. In Sec.~\ref{sec:5} we summarize our findings and address potential future work.

\section{Weak-field Non-local infinite-derivative gravity}
\label{sec:2}
Let us work in Cartesian coordinates $x{}^\mu = (t,x^i)$ and $x^i = (x,y,z)$ such that the Minkowski metric takes the form
\begin{align}
\dd s^2 = \eta{}_{\mu\nu} \dd x{}^\mu \dd x{}^\nu = -\dd t^2 + \dd x^2 + \dd y^2 + \dd z^2 \, .
\end{align}
Moreover, let us parametrize a perturbation $h{}_{\mu\nu}$ such that the full metric is
\begin{align}
g{}_{\mu\nu} = \eta{}_{\mu\nu} + h{}_{\mu\nu} \, , \quad g{}^{\mu\nu} = \eta{}^{\mu\nu} - h{}^{\mu\nu} \, .
\end{align}
For later convenience we also define the trace of the metric perturbation,
\begin{align}
h = \eta{}^{\mu\nu}h{}_{\mu\nu} \, .
\end{align}
Last, let us define the totally antisymmetric tensor $\epsilon{}_{\mu\nu\rho\sigma}$ as the volume element on Minkowski spacetime. To linear order, the spacetime curvature is
\begin{align}
\begin{split}
\label{eq:curvatures}
R{}_{\mu\nu\rho\sigma} &= \partial{}_\nu \partial{}_{[\rho} h{}_{\sigma]\mu} - \partial{}_\mu \partial{}_{[\rho} h{}_{\sigma]\nu} \, , \\
R{}_{\mu\nu} &= \eta{}^{\rho\sigma}R{}_{\mu\rho\nu\sigma} \\
&= \partial{}_\rho \partial{}_{(\mu} h{}_{\nu)}{}^\rho - \frac12 \left( \partial{}_\mu\partial{}_\nu h + \Box h{}_{\mu\nu} \right) \, , \\
R &= \eta{}^{\mu\nu}R{}_{\mu\nu} = \partial{}_\rho\partial{}_\sigma h{}^{\rho\sigma} - \Box h \, ,
\end{split}
\end{align}
where we denoted the d'Alembert and Laplace operators
\begin{align}
\Box = \eta{}^{\mu\nu}\partial{}_\mu\partial{}_\nu = -\partial_t^2 + \lap \, , \quad \lap = \partial_x^2 + \partial_y^2 + \partial_z^2 \, .
\end{align}
With the geometric setup in place, let us now study the weak-field model of non-local infinite-derivative gravity.

\subsection{Field equations}
The linearized field equations of non-local infinite-derivative gravity for a small perturbation $h{}_{\mu\nu}$ around Minkowski spacetime can be derived from the following action:
\begin{align}
\label{S}
S &= \frac{1}{2\kappa} \int \dd^D x \, \Big( \frac12 h{}^{\mu\nu} a(\Box) \Box h{}_{\mu\nu} - h{}^{\mu\nu} a(\Box) \partial{}_\mu \partial{}_\alpha h{}^\alpha{}_\nu \nonumber \\
&\hspace{57pt} + h{}^{\mu\nu} c(\Box) \partial{}_\mu \partial{}_\nu h - \frac12 h c(\Box) \Box h \\
&\hspace{57pt} + \frac12 h{}^{\mu\nu} \frac{a(\Box) - c(\Box)}{\Box} \partial{}_\mu \partial{}_\nu \partial{}_\alpha \partial{}_\beta h{}^{\alpha\beta} \Big) \, . \nonumber
\end{align}
They take the form
\begin{align}
\begin{split}
&a(\Box)\left[ \Box h{}_{\mu\nu} - 2 \partial_\rho \partial_{(\mu} h{}_{\nu)}{}^\rho \right] \\
&+ c(\Box)\left[ \eta{}_{\mu\nu}\left( \partial_\rho\partial_\sigma h{}^{\rho\sigma} - \Box h \right) + \partial{}_\mu \partial{}_\nu h \right] \\
&+ \frac{a(\Box)-c(\Box)}{\Box} \partial{}_\mu\partial{}_\nu\partial{}_\rho\partial{}_\sigma h{}^{\rho\sigma} = -2\kappa T{}_{\mu\nu} \, ,
\end{split}
\end{align}
where $\kappa = 8\pi G$ stands for Einstein's gravitational constant, and parentheses denote symmetrization,
\begin{align}
\partial{}_{(\mu} h{}_{\nu)\alpha} = \frac12 \left( \partial{}_\mu h{}_{\nu\alpha} + \partial{}_\nu h{}_{\mu\alpha} \right) \, .
\end{align}
One may verify that the field equations are consistent with $\partial{}^\mu T{}_{\mu\nu} = 0$. The functions $a(\Box)$ and $c(\Box)$ are called \emph{form factors} and parametrize the non-locality of the field equations. They are subject to the constraint
\begin{align}
a(0) = c(0) = 1 \, ,
\end{align}
which guarantees a proper Newtonian limit.

\subsection{Ricci curvature}
Using Eq.~\eqref{eq:curvatures} the field equations can be recast in terms of the Ricci curvature tensor as follows:
\begin{align}
\label{eq:eom-ac}
a(\Box) R{}_{\mu\nu} - \frac12\eta{}_{\mu\nu}c(\Box)R - \frac{a(\Box) - c(\Box)}{2\Box}\partial{}_\mu \partial{}_\nu R = \kappa T{}_{\mu\nu} \, .
\end{align}
Note that $\partial{}^\mu T{}_{\mu\nu} = 0$ implies that
\begin{align}
a(\Box)\partial{}^\mu \left( R{}_{\mu\nu} - \frac12 R \eta{}_{\mu\nu} \right) = 0 \, .
\end{align}
This corresponds to the usual contracted Bianchi identity for the Einstein tensor (in the weak-field limit), since in infinite-derivative non-local field theories discussed here we assume that the form factors are strictly non-vanishing such that they can be inverted.

For general $a(\Box)$ and $c(\Box)$ the field equation \eqref{eq:eom-ac} is not algebraic in the Ricci tensor, unlike in general relativity. In momentum space, however, it is possible to express the Ricci tensor via the energy-momentum tensor directly,
\begin{align}
\begin{split}
R{}_{\mu\nu} = \left[ \frac{1}{a_k}\delta{}^\alpha_\mu \delta{}^\beta_\nu + \frac{c_k}{a^2_k - 3a_k c_k}\eta{}_{\mu\nu}\eta{}^{\alpha\beta} \right. \\
+ \left. \frac{a_k-c_k}{a^2_k - 3a_k c_k} \frac{k{}_\mu k{}_\nu}{k^2} \eta{}^{\alpha\beta} \right] \kappa T{}_{\alpha\beta} \, ,
\end{split}
\end{align}
where we defined $a_k = a(-k^2)$ and $c_k = c(-k^2)$ for convenience, and $k^2$ denotes the square of the 4-momentum.

This implies that even at the linearized level, the interpretation of the Ricci curvature as the ``matter curvature'' is no longer valid in non-local theories of the above type. In particular, the above considerations also show that Ricci flat spacetimes, $R{}_{\mu\nu}=0$, are always vacuum spacetimes, $T{}_{\mu\nu} = 0$, but the converse is no longer true: it appears possible to construct vacuum spacetimes that have non-vanishing Ricci curvature.

From now on we shall focus on a special class of non-local theories where
\begin{align}
a(\Box) = c(\Box) \, .
\end{align}
Then the field equations \eqref{eq:eom-ac} simplify to
\begin{align}
\label{eq:eom-a}
a(\Box) \left( R{}_{\mu\nu} - \frac12 \eta{}_{\mu\nu} R \right) = \kappa T{}_{\mu\nu} \, ,
\end{align}
such that the Ricci curvature can be expressed as
\begin{align}
R{}_{\mu\nu} = a^{-1}(\Box)\left( T{}_{\mu\nu} - \frac12 \eta{}_{\mu\nu} T \right) \, .
\end{align}
Recall that in the local theory one has $a=1$ and hence the Ricci tensor and the energy-momentum tensor are linked \emph{algebraically.} In non-local theories, even at the linear level, this is no longer the case. The inverse operator $a^{-1}(\Box)$ always exists in non-local theories of this class since $a(\Box)$ has no zeroes. In the literature it has been shown that this inverse operator can act as a smearing operator on sharply localized objects, mostly in the static case but also in the time-dependent case \cite{Frolov:2016xhq,Kolar:2021oba}.

This allows for the tentative interpretation of the Ricci curvature as the ``smeared out matter curvature'' in this class of non-local theories. Moreover, this emphasizes the special role of the Weyl curvature in this class of theories as the only part of curvature that is not directly specified by the field equations. In this linear setting, the Weyl tensor can be written as
\begin{align}
\begin{split}
C{}_{\mu\nu\rho\sigma} = R{}_{\mu\nu\rho\sigma} &- \eta{}_{\mu[\rho} R{}_{\sigma]\nu} + \eta{}_{\nu[\rho} R{}_{\sigma]\mu} \\
&+ \frac13 R \eta{}_{\mu[\rho}\eta{}_{\sigma]\nu} \, ,
\end{split}
\end{align}
where square brackets denote antisymmetrization,
\begin{align}
\eta{}_{\mu[\rho}\eta{}_{\sigma]\nu} = \frac12 \left( \eta{}_{\mu\rho}\eta{}_{\sigma\nu} - \eta{}_{\mu\sigma}\eta{}_{\rho\nu} \right) \, .
\end{align}
The Weyl tensor can hence be interpreted as the difference between the full Riemann curvature and the smeared out matter curvature, and for that reason we shall refer to the Weyl tensor as the ``vacuum curvature.''

\subsection{$\mathrm{GF_N}$ model for non-local theories}
In what follows we will focus our considerations on so-called $\mathrm{GF_N}$ theory wherein
\begin{align}
a(\Box) = c(\Box) = \exp\left[ (-\ell^2\Box)^N \right] \, , \quad N \in \mathbb{N} \, ,
\end{align}
and $\ell>0$ denotes the \emph{scale of non-locality}. Clearly this form factor satisfies $a(0) = 1$, which also guarantees that one recovers the local theory in the limit $\ell\rightarrow 0$. In the time-independent case, which we study in this paper, this simplifies further to
\begin{align}
a(\lap) = \exp\left[ (-\ell^2\lap)^N \right] \, , \quad N \in \mathbb{N} \, ,
\end{align}
with the final form of the field equations
\begin{align}
\label{eq:eom-gfn-static}
\exp\left[ (-\ell^2\lap)^N \right] \left( R{}_{\mu\nu} - \frac12 \eta{}_{\mu\nu} R \right) = \kappa T{}_{\mu\nu} \, .
\end{align}
It is well known that in this case the field equations can be interpreted as the local Einstein equations with a smeared matter source,
\begin{align}
\begin{split}
 R{}_{\mu\nu} - \frac12 \eta{}_{\mu\nu} R  = \kappa T{}^\text{eff}_{\mu\nu} \, , \\[5pt]
T{}^\text{eff}_{\mu\nu} \equiv \exp\left[ -(-\ell^2\lap)^N \right] T{}_{\mu\nu} \, .
\end{split}
\end{align}
As has been shown in the literature, if $T{}_{\mu\nu}$ describes a sharply concentrated matter distribution, then the effective energy-momentum tensor is smeared out \cite{Giacchini:2018wlf}. In order to formalize this notion somewhat, as well as for later convenience, let us introduce the concept of a smeared $\delta$ function and a smeared Heaviside function as follows:
\begin{align}
\begin{split}
\delta{}^{(d)}_\ell(\ts{x}) &\equiv \exp\left[ -(-\ell^2\lap)^N \right] \delta{}^{(d)}(\ts{x}) \, , \\
\theta_\ell(x) &\equiv \exp\left[ -(-\ell^2\lap)^N \right] \theta(x) \, .
\end{split}
\end{align}
Note that these functions are related via
\begin{align}
\label{eq:delta-theta-relation}
\partial_x \theta_\ell(x) = \delta{}^{(1)}_\ell(x) \, ,
\end{align}
which follows from the formal identity $\partial_x \theta(x) = \delta^{(1)}(x)$ which can be verified in the distributional sense within an integral. In the limiting case of $\ell\rightarrow 0$ one recovers
\begin{align}
\label{eq:delta-theta-l=0}
\lim\limits_{\ell\rightarrow 0} \delta{}^{(d)}_\ell(\ts{x}) = \delta{}^{(d)}(\ts{x}) \, , \quad \lim\limits_{\ell\rightarrow 0} \theta{}_\ell(x) = \theta(x)
\end{align}
In the simplest case of $N=1$ one finds the expressions
\begin{align}
\begin{split}
\delta{}^{(d)}_\ell(\ts{x}) &= \frac{1}{(4\pi\ell^2)^{d/2}} e^{-\ts{x}^2/(4\ell^2)} \, , \\
\theta{}_\ell(x) &= \frac12 \left[ 1 + \text{erf}\left(\frac{x}{2\ell}\right) \right] \, ,
\end{split}
\end{align}
and one may verify that they satisfy Eqs.~\eqref{eq:delta-theta-relation} and \eqref{eq:delta-theta-l=0}. Let us mention that this smeared $\delta$-function appears in the definition of static non-local Green functions,
\begin{align}
\begin{split}
a(\lap)\lap\mathcal{G}_d(\ts{x}) &= -\delta{}^{(d)}(\ts{x}) \, , \\
\Leftrightarrow \quad \lap \mathcal{G}_d(\ts{x}) &= -\delta{}^{(d)}_\ell(\ts{x}) \, . 
\end{split}
\end{align}
Due to spherical symmetry $\mathcal{G}_d(\ts{x}-\ts{y})$ is a function of $r =|\ts{x}-\ts{y}|$ and hence in what follows we may abbreviate $\mathcal{G}_d(\ts{x}-\ts{y}) = \mathcal{G}_d(r)$. Last, let us note that the static Green functions are related via \cite{Frolov:2015usa}
\begin{align}
\label{eq:g-recursion}
\mathcal{G}_{d+2}(r) = -\frac{1}{2\pi r} \frac{\partial\mathcal{G}_d(r)}{\partial r} \, .
\end{align}
This allows a successive construction of non-local static Green functions from just two ``seed functions,'' and for a more in-depth reference on non-local spatial Green functions we refer to Ref.~\cite{Boos:2018bxf}. In the simplest case of $N=1$ a sufficient set of seed functions is
\begin{align}
\mathcal{G}_3(r) &= \frac{1}{4\pi r}\text{erf}\left(\frac{r}{2\ell}\right) \, , \\
\mathcal{G}_4(r) &= \frac{1}{4\pi^2r^2} \left[ 1 - e^{-r^2/(4\ell^2)} \right] \, .
\end{align}

\section{Gravitoelectromagnetic quantities and stationary weak-field metrics}
\label{sec:3}
In the present paper we are interested in gravitomagnetic properties of stationary geometries, which are defined by the presence of a timelike Killing vector $\ts{\xi} = \partial_t$ such that
\begin{align}
\label{eq:killing}
\mathcal{L}_\ts{\xi} h{}_{\mu\nu} = 0 \, .
\end{align}
Using this Killing vector we may define the \emph{electric} and \emph{magnetic} part of the Weyl tensor as follows \cite{Bel:1962,Bel:2000,Stephani:2003,Boos:2014hua}:
\begin{align}
\begin{split}
E{}_{ij} &= C{}_{\mu i\nu j}\xi{}^\mu\xi{}^\nu = C{}_{titj} \, , \\
B{}_{ij} &= \frac12 \epsilon{}_{\mu i\rho\sigma} C{}^{\rho\sigma}{}_{\nu j}\xi{}^\mu\xi{}^\nu = \frac12 \epsilon{}_{t i\rho\sigma} C{}^{\rho\sigma}{}_{tj} \, .
\end{split}
\end{align}
It follows from the antisymmetry in the pairs of indices of the Weyl tensor and the $\epsilon$-symbol that these tensors have no timelike components. Moreover, by the fundamental symmetry properties of the Weyl tensor these tensors are symmetric and tracefree,
\begin{align}
E{}_{[ij]} = B{}_{[ij]} = 0 \, , \quad \eta{}^{ij} E{}_{ij} = \eta{}^{ij} B{}_{ij} = 0 \, .
\end{align}
Therefore, they each encompass five independent components which encode the ten independent tensorial components of the four-dimensional Weyl tensor.

Let us make the stationary ansatz
\begin{align}
h{}_{\mu\nu}\dd x{}^\mu \dd x{}^\nu = \phi \dd t^2 + 2 A_i \dd x{}^i \dd t + h{}_{ij}\dd x{}^i\dd x{}^j \, ,
\end{align}
where $\partial_t \phi = \partial_t A_i = \partial_t h{}_{ij} = 0$ due to Eq.~\eqref{eq:killing}. It is also useful to define the quantities
\begin{align}
\epsilon{}_{ijk} = \epsilon{}_{tijk} \, , \quad F{}_{ij} = \partial_i A{}_j - \partial{}_j A{}_i \, , \quad \bar{R} = \delta{}^{ij}R{}_{ij} \, .
\end{align}
Then, the electric and magnetic parts of the Weyl tensor can be written as
\begin{align}
\label{eq:electric}
E{}_{ij} &= -\frac12\left( \partial_i\partial_j - \frac13\delta{}_{ij} \lap \right)\phi + \frac12\left( R{}_{ij} - \frac13\bar{R}\delta{}_{ij} \right) \, , \\
\label{eq:magnetic}
B{}_{ij} &= \frac14 \left( \partial_j \epsilon{}_i{}^{kl} F{}_{kl} - \epsilon{}_{ij}{}^l \partial{}^m F{}_{ml} \right) = \frac14 \partial{}_{(i} \epsilon{}_{j)}{}^{kl} F{}_{kl} \, .
\end{align}
Clearly, $E{}_{ij}$ is symmetric and tracefree. $B{}_{ij}$ is also tracefree since $\partial{}_{[i} F{}_{jk]} = 0$ by construction, and it is symmetric because its antisymmetric part vanishes:
\begin{align}
8B{}_{[ij]} &= \left( \partial{}_j \epsilon{}_i{}^{kl} - \partial{}_i \epsilon{}_j{}^{kl} - 2 \epsilon{}_{ij}{}^l \partial{}^k \right) F{}_{kl} \nonumber \\
&= \left( \partial{}_j \epsilon{}_i{}^{kl} - \partial{}_i \epsilon{}_j{}^{kl} - 2 \epsilon{}_{ij}{}^a \partial{}^b \delta{}^l_{[a} \delta{}^k_{b]} \right) F{}_{kl} \\
&= \left( \partial{}_j \epsilon{}_i{}^{kl} - \partial{}_i \epsilon{}_j{}^{kl} - 2 \epsilon{}_{ij}{}^a \partial{}^b \epsilon{}_{cab}\epsilon{}^{c\,lk} \right) F{}_{kl} = 0 \, , \nonumber
\end{align}
where we have employed the three-dimensional identity $\epsilon{}_{cab}\epsilon{}^{cij} = +\delta{}^i_{[a} \delta{}^j_{b]}$. Note that in case of spherical symmetry one has $h{}_{ij} = \psi\delta{}_{ij}$ and hence one can further simplify the structure of $E{}_{ij}$. One finds
\begin{align}
\label{eq:electric-2}
E{}_{ij} &= -\frac14\left( \partial_i\partial_j - \frac13\delta{}_{ij} \lap \right)(\phi+\psi) \, .
\end{align}
Suppose now that one calculates $E{}_{ij}$ and $B{}_{ij}$ for a Weyl tensor $C{}_{\mu\nu\rho\sigma}$. Then one can perform the duality rotation
\begin{align}
\label{eq:duality-transformation}
\tilde{C}{}_{\mu\nu\rho\sigma} = \frac12 \epsilon{}_{\mu\nu}{}^{\alpha\beta} C{}_{\alpha\beta\rho\sigma} \,, 
\end{align}
which maps the Weyl tensor into its left dual. Calculating the electric and magnetic pieces for this left dual of the Weyl tensor one finds
\begin{align}
\tilde{E}_{ij} &\equiv \tilde{C}{}_{titj} = B{}_{ij} \, , \\
\tilde{B}_{ij} &\equiv \frac12 \epsilon{}_{ti\rho\sigma} \tilde{C}{}^{\rho\sigma}{}_{tj} = -E{}_{ij} \, ,
\end{align}
which follows from the four-dimensional relation\footnote{This is the tensorial equivalent of the relation $\star\star=-1$ one encounters for the Hodge dual acting on differential forms on Lorentzian manifolds. It gives rise to an \emph{almost complex structure} and allows the notion of duality.}
\begin{align}
\epsilon{}_{\mu\nu\alpha\beta}\epsilon{}^{\alpha\beta\rho\sigma} = -2\delta{}^\rho_{[\mu}\delta{}^\sigma_{\nu]} \, .
\end{align}
This implies that, up to a sign, a duality transformation \eqref{eq:duality-transformation} maps gravitoelectric and gravitomagnetic quantities into each other.\footnote{In general relativity, this transformation maps the mass to the NUT parameter, and the angular momentum to the rotational parameter of the NUT solution \cite{Boos:2014hua}.}

\subsection{Schwarzschild and NUT solutions}
Let us now apply this formalism to study the GEM duality properties of weak-field solutions in non-local infinite-derivative gravity. In what follows we will consider the gravitational fields of a point particle (``Schwarzschild solution'') and that of a spinning semi-infinite string (``NUT solution''). In the derivation we assume the Lorenz gauge $\partial{}_\mu h{}^\mu{}_\nu = \tfrac12 \partial{}_\nu h$.

\subsubsection{Schwarzschild solution}
The weak-field Schwarzschild geometry is sourced by the distributional energy-momentum tensor
\begin{align}
T{}^\textsc{Schw}_{\mu\nu} = m \, \delta{}^t_\mu \delta{}^t_\nu \, \delta{}^{(3)}(\ts{x}) \, ,
\end{align}
which describes a static particle of mass $m>0$ at rest in the coordinate origin. Since the energy-momentum tensor is proportional to a three-dimensional $\delta$-function, the solution of Eq.~\eqref{eq:eom-gfn-static} is proportional to the non-local three-dimensional static Green function. For $N=1$ it takes the form \cite{Modesto:2011kw,Biswas:2011ar,Edholm:2016hbt,Boos:2018bxf}
\begin{align}
\begin{split}
& h{}^\textsc{Schw}_{\mu\nu}\dd x{}^\mu\dd x{}^\nu = \phi \dd t^2 + \psi \left( \dd x^2 + \dd y^2 + \dd z^2 \right) \, , \\
& \phi = \psi = \kappa m\mathcal{G}_3(r) = \frac{2Gm}{r}\text{erf}\left(\frac{r}{2\ell}\right) \, .
\end{split}
\end{align}
As has been discussed elsewhere in great detail, this solution is manifestly regular at $r=0$ and one asymptotically recovers the weak-field Schwarzschild solution of linearized general relativity as $r/\ell \rightarrow \infty$. Since the solution is given by the non-local Green function directly, it can readily be generalized to different $\mathrm{GF_N}$ theories.

\subsubsection{NUT solution}
The weak-field NUT solution, in its massless limit, is sourced by the following energy-momentum tensor:
\begin{align}
\label{eq:tmunu-nut}
T{}^\textsc{NUT}_{\mu\nu} &= - \delta{}^t_{(\mu} \delta{}^i_{\nu)} \, n{}_i{}^j \partial_j \, \delta{}^{(1)}(x)\delta{}^{(1)}(y)\theta(z) \, ,
\end{align}
where $n_{ij} = -n{}_{ji}$ is an antisymmetric tensor with
\begin{align}
n \equiv n{}_{xy} = - n_{yx} \, .
\end{align}
The solution of Eq.~\eqref{eq:eom-gfn-static} can be found analytically in the case of $N=1$ \cite{Kolar:2020bpo}. Here we rewrite it in terms of the smeared $\delta$-function and Heaviside function as follows:
\begin{align}
\begin{split}
\label{eq:nut-nonlocal}
& h{}^\textsc{NUT}_{\mu\nu}\dd x{}^\mu \dd x{}^\nu = 2 A_x \dd x \dd t + 2 A_y \dd y \dd t \, , \\
& A_i = \kappa n \frac{\epsilon{}_{ijk}x^j L^k}{\rho^2} V \, , \quad L^\mu = \delta{}^\mu_z \, , \\
& V = \frac{1}{4\pi} + z \mathcal{G}_3(r) - 2\ell^2\theta_\ell(z) \delta^{(2)}_\ell(\rho) \, , 
\end{split}
\end{align}
The third term in $V$ is interesting since it corresponds to a smeared positive $z$-axis; it vanishes identically in the local limit due to the $\ell^2$-prefactor. The compact and universal form of this solution suggests that it may be possible to construct this metric for other $\mathrm{GF_N}$ theories as well.\footnote{Formally it is possible to derive the NUT solution for any $\mathrm{GF_N}$ theory, see appendix \ref{app:NUT-derivation}.} The metric reduces to the general relativistic expression as $\rho/\ell\rightarrow\infty$, and one recovers the previously found metric of a slowly spinning string as $z\rightarrow+\infty$ \cite{Boos:2020kgj}.

\section{Gravitoelectromagnetic Schwarzschild--NUT dualities}
\label{sec:4}
As we have shown above, in non-local $\mathrm{GF_N}$ theories the Ricci tensor can be interpreted as a smeared matter curvature. Its tensorial structure is hence dictated by those of the energy-momentum tensor. In analogy to the local case we hence study the duality properties of the Weyl tensor alone. This step sets $\mathrm{GF_N}$ theories apart from (i) higher-derivative theories (even at the linear level) as well as (ii) non-local models at the non-linear level, where such an interpretation of the Ricci curvature is in general not possible.

\subsection{Broken duality}
With the weak-field solutions at our disposal, we can now evaluate the electric and magnetic parts of their respective Weyl tensors via Eq.~\eqref{eq:electric-2} and \eqref{eq:magnetic} and find
\begin{align}
E^\textsc{Schw}_{ij} &= -\frac{\kappa m}{2}\left[ \partial_i\partial_j \mathcal{G}_3(r) + \frac13 \delta{}_{ij} \delta{}^{(3)}_\ell(r) \right] \, , \\
B^\text{NUT}_{ij} &= +\frac{\kappa n}{2} \Big[ \delta{}_{ij}L^k\partial_k - (3+x^k\partial_k)L_{(i} \partial{}_{j)} \nonumber \\
&\hspace{40pt}+ x{}_{(i} \partial{}_{j)} L^k\partial_k \Big] \frac{V}{\rho^2} \, , \\
B^\textsc{Schw}_{ij} &= E^\text{NUT}_{ij} = 0 \, .
\end{align}
One may verify that these tensors are indeed tracefree. The difference of the electric Schwarzschild part and the magnetic NUT part for $n \rightarrow m$ is then
\begin{align}
\begin{split}
\label{eq:xi}
\Xi_{ij} &\equiv E^\textsc{Schw}_{ij}(m) - B^\text{NUT}_{ij}(n\rightarrow m) \\
&= \frac{\kappa m}{2}\left[ L_{(i} \partial_{j)} \theta_\ell(z) - \frac13\delta{}_{ij} \delta{}^{(1)}_\ell(z) \right] \delta{}^{(2)}_\ell(\rho) \not= 0 \, .
\end{split}
\end{align}
If the duality was exact, then one would have $\Xi{}_{ij} = 0$. Because it does not vanish, the GEM duality between the Schwarzschild and massless NUT solution is \emph{broken} at the linear level in the non-local theory.

In the local theory, however, the situation is different. Utilizing the relations \eqref{eq:delta-theta-l=0} in the limiting case of $\ell\rightarrow 0$ one finds instead
\begin{align}
\label{eq:duality-local}
\Xi_{ij}^{\ell\rightarrow 0} &= \frac{\kappa m}{2}\left[ L_{(i} \partial_{j)} \theta(z) - \frac13\delta{}_{ij} \delta(z) \right] \delta{}^{(2)}(\rho) \, ,
\end{align}
which is a distributional quantity that is non-vanishing on the positive $z$-axis. This corresponds to the sometimes overlooked fact that in weak-field general relativity the Schwarzschild and NUT solution are only dual to each other away from the positive $z$-axis, as was pointed out some time ago by Argurio and Dehouck \cite{Argurio:2009xr}.

This calculation justifies the interpretation of the scale of non-locality as a regulator, since the distributional quantities only appear in the local limit $\ell\rightarrow 0$. Hence, even if physics turns out to be ultimately local, ``non-local regularization'' may simply serve as a tool.

\subsection{Exact duality}
As just seen, in weak-field general relativity the GEM quantities exhibit distributional character on the positive $z$-axis. The study is hence mathematically more involved since, in principle, one would be required to employ distributional calculus to make sense of derivatives of distributions as encountered in Eq.~\eqref{eq:duality-local}. However, in the non-local theory this is not the case, and all functions encountered are smooth and differentiable for finite $\ell>0$. At any rate, in both setups there \emph{is no exact duality.}

In the local weak-field theory, Bunster \textit{et al.} \cite{Bunster:2006rt} propose a modified set of  gravitational equations that is manifestly invariant under duality transformations similar to \eqref{eq:duality-transformation}, albeit applied to the full Riemann tensor,
\begin{align}
\tilde{R}{}_{\mu\nu\rho\sigma} = \frac12 \epsilon{}_{\mu\nu}{}^{\alpha\beta} R{}_{\alpha\beta\rho\sigma} \, .
\end{align}
Let us call this model the ``BCHP model'' after its inventors. In spirit, this is similar to the inclusion of magnetic monopoles into the Maxwell equations; see Edelen \cite{Edelen:1978} and references therein. Within this BCHP model, as Arguiro and Dehouck demonstrate \cite{Argurio:2009xr}, the weak-field Schwarzschild--NUT duality becomes exact everywhere, including the positive $z$-axis.
Here we would like to extend this conclusion to our non-local $\mathrm{GF_1}$ model. This step is non-trivial since the GEM duality is manifestly violated by non-locality.

In the BCHP model, just as in general relativity, the fundamental variable is the metric tensor. There are, however, \emph{two} sources of gravity. The energy-momentum tensor $T{}_{\mu\nu}$ as well as an additional symmetric tensor $\Theta{}_{\mu\nu}$ which may be viewed as a gravitomagnetic monopole source. The gravitational equations take the form
\begin{align}
\label{eq:bchp-1}
G{}_{\mu\nu} &= \kappa T{}_{\mu\nu} \, , \\
\label{eq:bchp-2}
3R{}_{\mu[\nu\alpha\beta]} &= -\kappa \epsilon{}_{\nu\alpha\beta\gamma} \Theta{}^\gamma{}_\mu \, , \\
\label{eq:bchp-3}
R{}_{\mu\nu[\alpha\beta,\gamma]} &= 0 \, .
\end{align}
Since they are dual under the transformation $(R,\tilde{R},T,\Theta) \rightarrow (\tilde{R},-R,\Theta,-T)$ one may also write
\begin{align}
\label{eq:bchp-4}
\tilde{G}{}_{\mu\nu} &= \kappa \Theta{}_{\mu\nu} \, , \\
\label{eq:bchp-5}
3\tilde{R}{}_{\mu[\nu\alpha\beta]} &= +\kappa \epsilon{}_{\nu\alpha\beta\gamma} T{}^\gamma{}_\mu \, , \\
\label{eq:bchp-6}
\tilde{R}{}_{\mu\nu[\alpha\beta,\gamma]} &= 0 \, .
\end{align}
Here, $\tilde{G}_{\mu\nu}$ denotes the Einstein tensor derived from the dual tensor $\tilde{R}_{\mu\nu\rho\sigma}$. However, note that in the above $R{}_{\mu\nu\rho\sigma}$ does not admit the interpretation as a Riemannian curvature tensor because it does not satisfy the algebraic Bianchi identity as per Eq.~\eqref{eq:bchp-2}.

In order to interpret $\Theta{}_{\mu\nu}$ as a proper source term, it should be conserved. This can be achieved by expressing it as a divergence of an auxiliary object $\Phi{}^{\mu\nu}{}_\rho$ such that
\begin{align}
\label{eq:relation-theta-phi}
\Theta{}^\mu{}_\nu = -\frac{1}{2\kappa}\partial{}_\alpha \Phi{}^{\alpha\mu}{}_\nu \, , \quad \Phi{}^{\mu\nu}{}_\rho = -\Phi{}^{\nu\mu}{}_\rho \, .
\end{align}
The antisymmetry of $\Phi{}^{\mu\nu}{}_\rho$ implies the conservation law $\partial{}_\mu \Theta{}^\mu{}_\nu = 0$. Then, the object $R{}_{\mu\nu\rho\sigma}$ is related to the curvature tensor (called $r_{\mu\nu\rho\sigma}$ in this section) via
\begin{align}
R{}_{\mu\nu\rho\sigma} &\equiv r{}_{\mu\nu\rho\sigma} + \delta R{}_{\mu\nu\rho\sigma} \, , \nonumber \\
\delta R{}_{\mu\nu\rho\sigma} &\equiv \frac14 \epsilon{}_{\mu\nu\alpha\beta} \left( \partial{}_\rho \bar{\Phi}{}^{\alpha\beta}{}_\sigma - \partial{}_\sigma \bar{\Phi}{}^{\alpha\beta}{}_\rho \right) \, , \\
\bar{\Phi}{}^{\mu\nu}{}_\rho &\equiv \Phi{}^{\mu\nu}{}_\rho + \frac12 \left( \delta{}^\mu_\rho \Phi{}^\nu - \delta{}^\nu_\rho \Phi{}^\mu \right) \, , \quad \Phi{}^\nu = \Phi{}^{\nu\alpha}{}_\alpha \, . \nonumber
\end{align}
Recall that $G{}_{\mu\nu}$ in Eq.~\eqref{eq:bchp-1} is the Einstein tensor calculated from $R{}_{\mu\nu\rho\sigma}$. For our present discussion we simply note that the curvature tensor is modified by the presence of a putative conserved $\Theta{}_{\mu\nu}$ monopole source. Just as the Schwarzschild solution is sourced by the energy-momentum tensor
\begin{align}
T{}_{\mu\nu} = m \, \delta{}^t_\mu \delta{}^t_\nu \delta{}^{(3)}(\ts{x}) \, ,
\end{align}
in the BCHP model the NUT solution is sourced by
\begin{align}
\label{eq:monopole-source}
\Theta{}_{\mu\nu} = n \, \delta{}^t_\mu \delta{}^t_\nu \delta{}^{(3)}(\ts{x}) \, .
\end{align}
In order to check whether this mathematical setup solves the duality problem, we may simply calculate the contribution of the additional curvature term $\delta R{}_{\mu\nu\rho\sigma}$ to the electromagnetic pieces of the Weyl tensor. To that end, the monopole source \eqref{eq:monopole-source} corresponds to
\begin{align}
\Phi{}^{zt}{}_t = -\Phi{}^{tz}{}_t = 2\kappa n \, \delta{}^{(1)}(x)\delta{}^{(1)}(y)\theta(z) \, .
\end{align}
Since the non-local $\mathrm{GF_N}$ theory, at the linear level, is equivalent to the local theory with smeared out sources, in what follows we consider the influence of the source
\begin{align}
\Theta{}^\text{eff}_{\mu\nu} = n \, \delta{}^t_\mu \delta{}^t_\nu \delta{}^{(3)}_\ell(\ts{x}) \, ,
\end{align}
mediated via
\begin{align}
\Phi{}^{\text{eff}~zt}{}_t{} = -\Phi{}^{\text{eff}~tz}{}_t = 2\kappa n \, \delta{}^{(1)}_\ell(x)\delta{}^{(1)}_\ell(y)\theta_\ell(z) \, .
\end{align}
The resulting contributions to the electric and magnetic parts of the Weyl tensor can be readily computed:
\begin{align}
\delta E{}_{ij} &\equiv \Pi{}_{ij}^{kl} \, \delta R{}_{tktl} = 0 \, , \\
\delta B{}_{ij} &\equiv \frac12 \Pi{}_{ij}^{kl} \epsilon{}_{tk\rho\sigma} \, \delta R{}^{\rho\sigma}{}_{tl} \nonumber \\
&= \frac{\kappa n}{2} \bigg[ L{}_{(i} \partial{}_{j)} \theta_\ell(z) - \frac13\delta{}_{ij} \delta{}^{(1)}_\ell(z) \bigg] \delta{}^{(2)}_\ell(\rho) \, ,
\end{align}
where we defined the projection operator
\begin{align}
\Pi{}_{ij}^{kl} = \delta{}^k_{(i} \delta{}^l_{j)} - \frac13 \eta{}_{ij}\eta{}^{kl} \, ,
\end{align}
which extracts the symmetric and traceless part of a rank-2 tensor.\footnote{We did not calculate the full Weyl tensor for the modified Riemann tensor $R{}_{\mu\nu\rho\sigma}$ since it violates the equality $R{}_{\mu[\nu\rho\sigma]}=0$ and has hence more irreducible pieces.} This result for $\delta B{}_{ij}$ precisely coincides with the discrepancy $\Xi{}_{ij}$ found in Eq.~\eqref{eq:xi} and thereby \emph{manifestly restores} the exact GEM duality.

The same is true for the local case, as already worked out by Argurio and Dehouck \cite{Argurio:2009xr}. We can recover their solution via the limiting procedure
\begin{align}
\delta B{}_{ij}^{\ell\rightarrow 0} = \frac{\kappa n}{2} \bigg[ L{}_{(i} \partial{}_{j)} \theta(z) - \frac13\delta{}_{ij} \delta{}^{(1)}(z) \bigg] \delta{}^{(2)}(\rho) \, ,
\end{align}
which is a distributional quantity  non-vanishing only on the positive $z$-axis. Let us emphasize that in our non-local $\mathrm{GF_1}$ model no such distributional quantities appear.

This construction shows that non-locality, at the linear level, exacerbates the violation of GEM duality into regions away from the positive $z$-axis. However, as we just demonstrated, it can be restored \emph{precisely} by the same procedure that is required in the local case.

\subsection{A non-local BCHP model}
Based on the successful application of the \emph{local} BCHP model to the weak-field sector with smeared sources, we would like to propose the following non-local generalization of the BCHP model:
\begin{align}
\label{eq:bchp-gf-1}
f(\lap) G{}_{\mu\nu} &= \kappa T{}_{\mu\nu} \, , \\
\label{eq:bchp-gf-2}
3f(\lap) R{}_{\mu[\nu\alpha\beta]} &= -\kappa \epsilon{}_{\nu\alpha\beta\gamma} \Theta{}^\gamma{}_\mu \, , \\
\label{eq:bchp-gf-3}
R{}_{\mu\nu[\alpha\beta,\gamma]} &= 0 \, .
\end{align}
Here, $f(\lap)$ is a non-local operator that satisfies $f(0) = 1$ and is formally given as a power series of the Laplace operator. Equivalently, due to their manifest GEM duality, we may write the field equations as
\begin{align}
\label{eq:bchp-gf-4}
f(\lap) \tilde{G}{}_{\mu\nu} &= \kappa \Theta{}_{\mu\nu} \, , \\
\label{eq:bchp-gf-5}
3f(\lap) \tilde{R}{}_{\mu[\nu\alpha\beta]} &= +\kappa \epsilon{}_{\nu\alpha\beta\gamma} T{}^\gamma{}_\mu \, , \\
\label{eq:bchp-gf-6}
\tilde{R}{}_{\mu\nu[\alpha\beta,\gamma]} &= 0 \, .
\end{align}
Based on our previous considerations, the following metric is a manifestly self-dual solution in this framework:
\begin{align}
\begin{split}
& h{}_{\mu\nu}\dd x{}^\mu \dd x{}^\nu = \phi \left( \dd t^2 + \dd x^2 + \dd y^2 + \dd z^2 \right) \\
&\hspace{81pt}+ 2 A_x \dd x \dd t + 2 A_y \dd y \dd t \, , \\
& \phi = \frac{2Gm}{r}\text{erf}\left(\frac{r}{2\ell}\right) \, , \quad A_i = \kappa n \frac{\epsilon{}_{ijk}x^j L^k}{\rho^2} V \, , \\
& V = \frac{1}{4\pi} + z \mathcal{G}_3(r) - 2\ell^2\theta_\ell(z) \delta^{(2)}_\ell(\rho) \, , \quad  \quad L^\mu = \delta{}^\mu_z \, .
\end{split}
\end{align}
It is sourced by the expressions
\begin{align}
T{}_{\mu\nu} = m \, \delta{}^t_\mu \delta{}^t_\nu \delta{}^{(3)}(\ts{x}) \, , \quad
\Theta{}_{\mu\nu} = n \, \delta{}^t_\mu \delta{}^t_\nu \delta{}^{(3)}(\ts{x}) \, .
\end{align}
Interestingly, the restoration of GEM duality did not require any change in the structure of the metric functions or the source terms, and has solely been accomplished by a modification of the field equations. The price to pay was the interpretation of $R{}_{\mu\nu\rho\sigma}$ as a curvature tensor: since it no longer satisfies the algebraic Bianchi identity, it may perhaps be regarded as a torsionful curvature tensor \cite{Hehl:1976kj}; see also Ref.~\cite{Kol:2020zth}.

Even though the explicit considerations of this paper are devoted to an understanding of the linearized Schwarzschild and NUT solutions, it is clear from the manifestly self-dual form of the non-local BCHP equations that similar relations hold for many other non-local solutions. In fact, two static solutions that are dual in the local BCHP model remain dual in the non-local extension.

\subsection{Harnessing duality structures}

In this last section we would like to briefly mention possible applications where the duality structures can be harnessed. To that end, recall that solutions with a given $\Theta{}_{\mu\nu}$-source can always be mapped into solutions of the regular Einstein equations with a $T{}_{\mu\nu}$-source. In other words, the modification term $\delta R{}_{\mu\nu\rho\sigma}$, as per Eq.~\eqref{eq:bchp-1}, can be moved to the right-hand side and viewed as a contribution to the energy-momentum tensor,
\begin{align}
\begin{split}
\delta T{}_{\mu\nu} &= -\frac{1}{4\kappa}\Big( ~ \partial{}_\nu \epsilon{}_{\mu\alpha\beta\gamma} \Phi{}^{\alpha\beta\gamma} - \epsilon{}_{\mu\alpha\beta\gamma} \partial{}^\alpha \Phi{}^{\beta\gamma}{}_\nu \\
&\hspace{35pt} + \epsilon{}_{\mu\nu\alpha\beta} \partial{}^\alpha \Phi{}^\beta - \eta{}_{\mu\nu} \epsilon{}_{\alpha\beta\gamma\delta} \partial{}^\alpha \Phi{}^{\beta\gamma\delta} \Big) \, .
\end{split}
\end{align}
Is this contribution always symmetric? The answer is yes, if and only if $\Theta{}_{\mu\nu}$ is symmetric, which we assume throughout in accordance with Ref.~\cite{Bunster:2006rt}. The easiest way to prove this is from considering the cyclic Bianchi identity \eqref{eq:bchp-2}, from which one may derive an antisymmetric part of the Riemann tensor
\begin{align}
\begin{split}
R{}_{\mu\nu\alpha\beta}-R{}_{\alpha\beta\mu\nu} &= \delta R{}_{\mu\nu\alpha\beta} - \delta R{}_{\alpha\beta\mu\nu} \\
&= -\frac{\kappa}{2}\big( \epsilon{}_{\mu\nu\alpha\lambda}\Theta{}^\lambda{}_\beta - \epsilon{}_{\mu\nu\beta\lambda}\Theta{}^\lambda{}_\alpha \\
&\hspace{22pt}- \epsilon{}_{\alpha\beta\mu\lambda}\Theta{}^\lambda{}_\nu + \epsilon{}_{\alpha\beta\nu\lambda}\Theta{}^\lambda{}_\mu \big) \, .
\end{split}
\end{align}
Note that the above expression vanishes for a pure Riemann tensor $r{}_{\mu\nu\alpha\beta}$, which is why this contribution is proportional to the gravitomagnetic source term $\Theta{}_{\mu\nu}$. It induces a potentially antisymmetric part to the Ricci tensor according to
\begin{align}
\delta R{}_{[\mu\alpha]} = \eta{}^{\nu\beta} ( \delta R{}_{\mu\nu\alpha\beta} - \delta R{}_{\alpha\beta\mu\nu} ) = -\kappa \epsilon{}_{\mu\alpha\gamma\delta}\Theta{}^{\gamma\delta} \, .
\end{align}
However, since $\Theta{}_{\mu\nu} = \Theta{}_{\nu\mu}$, this antisymmetric part of the Ricci tensor modification vanishes. This constitutes an important consistency check of the resulting effective Einstein equations.

We can use this duality structure as follows. Start with the energy-momentum tensor $T{}_{\mu\nu}$ of a seed metric, of which the solution to the non-local Einstein equations is known. Then, by means of the duality, set $\Theta{}_{\mu\nu} = T{}_{\mu\nu}$ and use the relations above to determine the resulting energy-momentum tensor from that choice. The solution of the resulting non-local Einstein equation will yield the dual solution for the original seed metric.

In the context of our previous example, we began with a point particle solution where $T{}_{\mu\nu} \sim \delta{}^t_\mu \delta{}^t_\nu \delta{}^{(3)}(\ts{x})$. The weak-field solution is the non-local Schwarzschild metric. Then, one may stipulate instead a monopole source of the same form, $\Theta{}_{\mu\nu} \sim \delta{}^t_\mu \delta{}^t_\nu \delta{}^{(3)}(\ts{x})$, which gives rise to non-vanishing components $\delta T{}_{ti}$ with $i=x,y$. Then, the resulting Einstein equations are solved by the massless NUT solution. Hence the interesting features of the BCHP model and its non-local extension therefore lie in the clever distribution of matter sources in the field equations, whereas the differential properties of the field equations remain essentially unchanged.

While a systematic survey of self-dual non-local solutions is beyond the scope of this paper we believe that the tools presented here serve as an ideal starting point for such inquiries.


\section{Conclusions}
\label{sec:5}
In this paper we have studied the fate of GEM duality for weak-field non-local gravity. As a testing ground, we considered the gravitational field of a point particle (Schwarzschild solution, ``gravitoelectric monopole'') and a semi-infinite spinning string (massless NUT solution, ``gravitomagnetic monopole''). In the case of linearized general relativity, these solutions are dual to each other everywhere except on the positive $z$-axis, where the duality is violated explicitly by distributional expressions. Since the realm of violation coincides with the location of matter sources, it may still be regarded as exact.

In this paper we showed that non-locality smears this violation of exact GEM duality to finite transverse distances away from the $z$-axis, the characteristic scale being the scale of non-locality $\ell$. In other words: non-locality \emph{spoils} any exact GEM duality.

Viewed from a different perspective, the existence of $\delta$-sources in general relativity has long been an active field of investigation; see the seminal work by Geroch and Traschen \cite{Geroch:1987qn}, or the more recent discussion by Pantoja and Rago \cite{Pantoja:2000cq}. Here we demonstrated that non-locality can serve as a regulator that turns distributional expressions ($\delta$-functions and derivatives thereof) into smooth functions. We emphasized this feature by introducing a notion of emergent $\delta$-functions and Heaviside functions. In the limiting case of $\ell\rightarrow 0$, we recover the results of linearized general relativity.

However, since the GEM duality is not exact even in linearized general relativity due to distributional quantities on the positive $z$-axis, Bunster \textit{et al.} \cite{Bunster:2006rt} developed a manifestly dual set of gravitational field equations that involves an additional gravitational source term. Applying their model to the non-local setup with smeared matter sources, we demonstrated that this procedure indeed \emph{solves the duality problem} in the class of non-local theories under consideration in this paper. In our calculations we relied heavily on the notion of effective $\delta$-functions, which in the mathematical literature are sometimes referred to as \emph{nascent $\delta$-functions}: these functions depend on the scale of non-locality $\ell>0$, and reduce to their usual behavior in the limiting case of $\ell\rightarrow 0$.

Last, guided by the successful adoption of the local gravitational model by Bunster \textit{et al.} to the non-local case, we extended their field equations to a non-local model by including infinite-derivative non-local form factors. We demonstrated that this non-local model maps dual solutions of the local theory into dual solutions of the non-local theory, which significantly extends the conclusions from the simple non-local Schwarzschild--NUT duality to far more general scenarios. Finally, we commented on how this self-duality structure of our non-local model can be employed to construct dual solutions to well-known non-local geometries.

Even though the considerations presented in this paper are only applicable to the weak-field regime, they present an important consistency check of non-local infinite-derivative gravity. In close proximity to matter sources one may expect that the full, non-linear non-local theory will lead to further modifications of GEM dualities, but we shall leave that discussion open for the future.


\section*{Acknowledgements}
J.B.\ would like to thank Friedrich W.~Hehl (Cologne) for bringing Ref.~\cite{Edelen:1978} to his attention, and is grateful for support by the National Science Foundation under grant PHY-181957. I.K. was supported by Netherlands Organization for Scientific Research (NWO) grant no. 680-91-119.

\appendix

\section{Derivation of the NUT solution}
\label{app:NUT-derivation}
The NUT solution for weak-field non-local gravity has been constructed via Laplace transform methods in Ref.~\cite{Kolar:2020bpo}. Here we would like to briefly delineate a possibly simpler derivation of the NUT solution that simultaneously extends to more general non-local theories of the $\mathrm{GF_N}$ type. Let us recall the energy-momentum tensor of the NUT source,
\begin{align}
\tag{\ref*{eq:tmunu-nut}}
T{}^\textsc{NUT}_{\mu\nu} &= - \delta{}^t_{(\mu} \delta{}^i_{\nu)} \, n{}_i{}^j \partial_j \, \delta{}^{(1)}(x)\delta{}^{(1)}(y)\theta(z) \, ,
\end{align}
with $n = n_{xy} = -n_{yx}$. Inserting the following ansatz into the stationary field equations \eqref{eq:eom-gfn-static},
\begin{align}
h{}^\textsc{NUT}_{\mu\nu}\dd x{}^\mu \dd x{}^\nu = 2 A_x \dd x \dd t + 2 A_y \dd y \dd t \, ,
\end{align}
and differentiating with respect to $z$ one finds
\begin{align}
e^{(-\ell^2\lap)^N} \lap A_i' = n{}_i{}^j \partial_j \, \delta{}^{(3)}(\ts{x}) \, , \quad A_i' \equiv \partial_z A_i \, .
\end{align}
This is solved by a rotating solution, recently discussed in Ref.~\cite{Boos:2020ccj}, taking the form
\begin{align}
& A_i' = - \kappa \, n{}_i{}^j \partial_j \mathcal{G}_3(r) \, , \quad r^2 = x^2+y^2+z^2 \, , \\
& e^{(-\ell^2\lap)^N} \lap \mathcal{G}_3(\ts{x}) = -\delta{}^{(3)}(\ts{x}) \, .
\end{align}
The form of $\mathcal{G}_3(r)$ is known for various $N$ and has been given in the literature, see e.g.~Ref.~\cite{Boos:2018bxf}. The final solution is hence obtained via integration over $z$,
\begin{align}
A_i = - \kappa \, n{}_i{}^j \partial_j \int\limits^z_{-\infty}\dd \tilde{z} \, \mathcal{G}_3\left( \sqrt{x^2+y^2+\tilde{z}^2} \right) \, .
\end{align}
In the simplest case for $N=1$ this integral can be performed analytically and one precisely recovers Eq.~\eqref{eq:nut-nonlocal}. Employing the recursion relation \eqref{eq:g-recursion} for non-local static Green functions one may write the equivalent
\begin{align}
A_i = 2\pi \kappa \, n{}_{ij}x^j \int\limits^z_{-\infty}\dd \tilde{z} \, \mathcal{G}_5\left( \sqrt{x^2+y^2+\tilde{z}^2} \right) \, .
\end{align}


\bibliography{references}

\end{document}